\documentclass[twocolumn,showpacs,prb,floatfix,superscriptaddress,citeautoscript,cite]{revtex4}
\usepackage{graphicx}
\usepackage{color}
\usepackage{amsmath}

\usepackage{booktabs}

\begin{document}

\title{Electronic and Magnetic Properties of 1T-TiSe$_{2}$ Nanoribbons}

\author{H. D. Ozaydin}
\email{hediyesengun@iyte.edu.tr}
\affiliation{Department of Physics, Izmir Institute of Technology, 35430 Izmir, Turkey}

\author{H. Sahin}
\affiliation{Department of Physics, University of Antwerp, Campus Groenenborgerlaan, 2020 Antwerp, Belgium}

\author{J. Kang}
\affiliation{Department of Physics, University of Antwerp, Campus Groenenborgerlaan, 2020 Antwerp, Belgium}

\author{F. M. Peeters}
\affiliation{Department of Physics, University of Antwerp, Campus Groenenborgerlaan, 2020 Antwerp, Belgium}

\author{R. T. Senger}
\email{tugrulsenger@iyte.edu.tr}
\affiliation{Department of Physics, Izmir Institute of Technology, 35430 Izmir, Turkey}

\date{\today}



\begin{abstract}

Motivated by the recent synthesis of single layer TiSe$_{2}$, we used state-of-the-art density functional theory 
calculations, to investigate the structural and electronic properties of zigzag and armchair-edged nanoribbons of this 
material. Our 
analysis reveals that, differing from ribbons of other ultra-thin materials such as graphene, TiSe$_{2}$  nanoribbons 
have some distinctive  properties. The electronic band gap of the nanoribbons decreases exponentially with the width 
and vanishes for ribbons wider than $20$ Angstroms. For ultranarrow zigzag-edged nanoribbons we find odd-even 
oscillations in the band gap width, although their band structures show similar 
features. Moreover, our detailed magnetic-ground-state  analysis reveals that zigzag and armchair edged ribbons 
have nonmagnetic ground states. Passivating the dangling bonds with hydrogen at the edges of the structures influences 
the band dispersion. Our  results shed light on the characteristic 
properties of T phase nanoribbons of similar crystal structures.

\end{abstract}

\maketitle

\section{Introduction}

Following the first experimental demonstration of graphene\cite{gr-1}, two-dimensional (2D) materials have attracted 
increasing attention both experimentally and theoretically. Especially transition metal dichalcogenides 
(TMDs)\cite{mx2-1,mx2-2,mx2-3} with chemical formula  MX$_2$ (where M is a transition metal atom and X is a chalcogen 
atom) have been a favored subject. There are also other stoichiometric forms of transition metal dichalcogenides such 
as titanium trisulfide (TiS$_3$) that can form monolayer crystals.\cite{fadil-tis3} TMDs have a special 2D layered 
structure. Their 
mono- and few-layered forms offer many opportunities for fundamental and technological 
research\cite{mx2-8,mx2-9,mx2-10} because of their 
exceptional  electronic, mechanical and optical properties\cite{mx2-4,mx2-5,mx2-6,mx2-7}. Furthermore, it is 
well known that various kinds of TMDs such as MoS$_{2}$, WS$_{2}$, MoSe$_{2}$, WSe$_{2}$, ReS$_{2}$, NbS$_2$, TiS$_2$, 
and TiSe$_2$ have been synthesized\cite{mx2-1,exp-ws,st-mose,sh-mose,hs-wse,hs-res,sh-res} and  studies have 
revealed that TMDs exhibit metallic, semimetallic, semiconducting, and even superconducting behavior with  different 
phases such as $1$H, $1$T and their distorted forms. 

The presence of exotic properties in $2$D materials, that stemmed from increasing quantum confinement effects, has also 
motivated researchers to further reduce their dimension and to investigate one-dimensional ($1$D) nanoribbons. 
In early studies it was shown that armchair and zigzag graphene nanoribbons (NRs) were semiconductors with  an energy 
gap 
decreasing with increasing ribbon width.\cite{han-GNR,abanin-GNR,sahin-GNR,veronica-gnr} In addition, zigzag graphene 
nanoribbons 
(ZGNRs) have ferromagnetically ordered edge states and can display half-metallic behavior when an external electric  
field is applied.\cite{son-gnr} Furthermore, motivated by the potential use of single layer MoS$_2$ in nanoscale 
optoelectronic devices, its nanoribbons  have been studied intensively.\cite{li-mos2nrsynt,mendez-mos2nr,pan-mos2nr}  
Armchair MoS$_2$ NRs are direct band gap semiconductors with a nonmagnetic ground state. Unlike GNRs their band gaps do 
not vary significantly with the ribbon width.\cite{sahin-mos2nr} However, zigzag MoS$_2$ NRs are ferromagnetic metals 
regardless of  their width and thickness.\cite{li-mos2nr}

Despite the comprehensive research on graphene and single layer TMDs, studies on the electronic properties of the group 
IVB TMDs in  the T phase, namely the two-dimensional $1$T-MX$_2$ structures, are sparse. Nevertheless, 
$1$T-TiSe$_2$\cite{salvo,fang,li,kusmartseva} is an extensively studied quasi-$2$D TMD, which has a charge density wave 
(CDW) state and in condensed matter physics transitions from superconductivity to charge  density wave phases has been 
shown to be very 
important\cite{sprcnd-1,sprcnd-2}. However, whether  $1$T-TiSe$_2$  is a semimetal or 
a semiconductor is still an open question\cite{rasch}. Since TiTe$_2$ is a semimetal with overlapping  valence and 
conduction bands\cite{tite2-1,tite2-2} and TiS$_2$ is 
a semiconductor with an indirect gap\cite{tis2-1,tis2-2}, it can be expected that the band gap of TiSe$_2$ is smaller 
or even 
nonexistent. Note that in the periodic table selenium is in between sulfur and tellurium, and also selenium is less 
electronegative than sulfur. Therefore, both experimental and theoretical techniques have been used to identify the 
semiconducting or 
semimetallic nature of 1T-TiSe$_2$\cite{pillo,calandra,hilbedrand,rosner}. Very recently, Peng et al.\cite{peng} 
grew TiSe$_2$ 
ultrathin films on a graphitized SiC(0001) substrate by  using molecular beam epitaxy (MBE). Their findings offer 
important insights into the nature of the charge density wave in TiSe$_2$, and paved the way for 
potential 
applications based on its collective electronic states\cite{peng}. The successful MBE growth of TiSe$_2$ ultrathin 
films 
down to monolayer thickness motivated us to investigate one-dimensional TiSe$_2$ nanoribbons because of its interesting 
electronic and physical properties that are essentially related  with its low dimensionality and effects due to quantum 
confinement. The main 
goal of this study is to find the characteristics of zigzag- and armchair-edged $1$T-TiSe$_2$ 
nanoribbons.

The paper is organized as follows. Details of the computational methodology are given 
in Sec. II. The calculated structural and electronic properties of single layer 1T-TiSe$_2$ are described in Sec. III. 
Then we analyze 1T-TiSe$_2$ nanoribbons and present results from spin-unpolarized and spin-polarized calculations in  
detail in Sec. IV. The last section, Sec. V, is devoted to the conclusion.

\begin{figure}
\includegraphics[width=8.5cm]{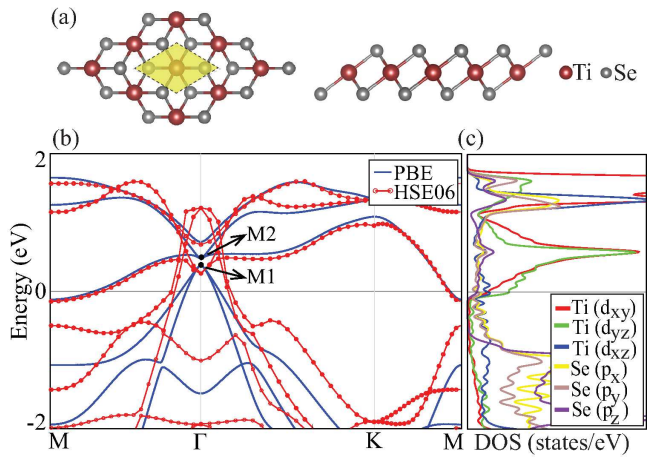}
\caption{\label{bandmonotise2}
(Color online) (a) Atomic structure of monolayer 1T-TiSe$_2$ with top and side views where the dashed yellow 
area denotes the unitcell of the monolayer,  and (b) the band structure calculated with PBE and HSE06, (c) 
partial 
density of states as calculated with PBE. Labels M$1$ and M$2$ are discussed in Fig. \ref{zigzagband-dec}.}
\end{figure}

\section{Computational Methodology}\label{comp}

The optimized structures and electronic properties of $1$T-TiSe$_2$ nanoribbons with desired edges (zigzag or armchair)  
reported here are based on first-principle calculations within the density functional theory (DFT) using the plane-wave 
projector-augmented wave (PAW) method\cite{paw} implemented in the Vienna \textit{ab-initio} simulation  package 
(VASP).\cite{vasp-1,vasp-2,vasp-3} The Perdew-Burke-Ernzerhof (PBE)\cite{pbe} form of the Generalized Gradient 
Approximation (GGA) were adopted to describe the electron exchange and correlation for both spin-polarized and 
spin-unpolarized cases.

In order to correct the PBE band structure for a monolayer of TiSe$_2$, we also used the Heyd-Scuseria-Ernzerhof 06 
(HSE06) 
functional\cite{hse1,hse2} which is known to give better electronic structure description that is close to experiments 
and produce accurate  band gaps. Since it improves the accuracy of standard band gaps, we determined HSE06 functional 
parameters as an enhanced fraction of the Hartree-Fock exchange $\alpha$ = 0.25 and screening $0.2$ \AA{}$^{-1}$. The 
kinetic energy cutoff for the plane-wave expansion was set to  $500$ eV where the  Brillouin Zone (BZ) was sampled with 
Monkhorst Pack (MP) by 7$\times$1$\times$1 k-point grids. For all band structure calculations, we used a 
75$\times$1$\times$1  $\Gamma$-centered k-point mesh. To avoid the  
interaction between  periodic images, we ensured a sufficient large supercell which is $20$\AA{} long perpendicular to 
the nanoribbon plane and with an edge-to-edge distance of at least $13$ \AA{}. At the same time, all the atoms in the 
supercell were 
fully relaxed during the geometry  optimization. The  convergence threshold for energy was chosen as 10$^{-5}$ eV and 
10$^{-4}$ eV/\AA{} for the force. The  charge distribution  on the atoms were calculated by using the Bader 
analysis.\cite{bader-1,bader-2}

Moreover, we investigated hydrogen saturated nanoribbons in order to study the edge stability. The hydrogen saturation 
was 
realized by adding one hydrogen atom to the edge of Ti and Se atoms for the zigzag nanoribbons, however for the 
armchair 
nanoribbons one hydrogen atom was added to the edge of Se atoms and two hydrogen atoms are added to the Ti atom. For 
the 
determination 
of the most favorable structure which means  the structure after hydrogenation, the binding 
energies were estimated from: E$_{B}$=E$_{T}$[NR]+nE$_{T}$[H]-E$_{T}$[NR+nH] where E$_{T}$[NR] is the total energy of 
the TiSe$_2$ nanoribbon, E$_{T}$[H] is the energy of the free hydrogen atom, E$_{T}$[NR+nH] is the total energy of the 
TiSe$_2$ nanoribbon saturated by hydrogen atoms, and n is the total number of saturated hydrogen atoms.

\section{Two-Dimensional Monolayer TiSe$_2$}

Before a comprehensive investigation of  TiSe$_2$ nanoribbons, we first present the atomic, electronic and magnetic 
properties of the TiSe$_2$ monolayer. Principally, layered structures of TMDs can form several different phases, $e.g.$ 
H and T, that result in diverse electronic properties. Monolayer  TiSe$_2$ has a hexagonal crystal  structure 
composed of three atom layers with a metal atom Ti layer sandwiched between two chalcogen Se layers. Here octahedral 
coordination of the metal atoms results in the $1$T structure as shown in Fig. \ref{bandmonotise2}(a).  Similar to 
graphite and 
graphene, in bulk TiSe$_2$ the monolayers are bound together through the interlayer van der Waals (vdW) interaction.  
The 
bond lengths are uniformly d$_{Ti-Se}$=$2.56$\AA{}, d$_{Se-Se}$=$3.72$\AA{},  where the angle between the Ti-Se bonds is
 $\theta_{Se-Ti-Se}$=$93.12^{\circ}$ and the optimized lattice  constant is $3.52$\AA{} from PBE calculation.

The PBE electronic band dispersion, shown in Fig. \ref{bandmonotise2}(b), shows that single layer TiSe$_2$ is a metal 
with a nonmagnetic ground state. In addition, the partial density of states (PDOS) reveals that while there is 
negligible contribution from the Se orbitals around the Fermi level (E$_{F}$), those bands are mainly composed of 
Ti-$3$d 
orbitals (d$_{z^2}$, d$_{xy}$, d$_{yz}$). At the same time, a Bader analysis indicates that each Ti atom gives 1.4 
electrons to the Se atoms  which means that 0.7 electrons are taken by one Se atom, hence this situation shows that the 
character of the bonding  is ionic. In contrast, the band structure of 1T TiS$_2$ is semiconducting. 
Usually the difference in chalcogen atoms affects the structural properties, but has little influence on the electronic 
properties. For instance single layers of  MoSe$_2$ and  MoS$_2$ are both direct band gap semiconductors.  However, a 
TiSe$_2$ sheet exhibits a metallic behavior with a low band crossing of the Fermi level, which is different from 
TiS$_2$.

To further examine  the electronic properties of 1T-TiSe$_2$, we also calculated the band structure with the HSE06 method  which is shown in Fig. \ref{bandmonotise2}(b). As can be seen the calculated bands below the Fermi level are shifted upward while above the Fermi level they are slightly shifted downward. At the same time, below the Fermi level the bands are decomposed but the bands above the Fermi level  almost overlap with those of the PBE 
result. In general, relative to the experimental values, band gaps of  semiconducting materials are underestimated by PBE. However, PBE+HSE06 provides better aggrement with the experimental values. Applying HSE06 corrections to metallic systems is not very common due to its computational cost, and no expected qualitative change in the band structures. 
Its effect is to introduce some shifts to the bands but the metallic character is preserved. For instance, single-layered VS$_2$ and T-MoS$_2$ are still found metallic with HSE06 correction.\cite{vs2,t-mos2}  Consequently, from both the PBE and HSE06 methods we may conclude that TiSe$_2$ is metallic.

\section{Nanoribbons of $1$T-TiSe$_2$ }

\subsection{Structural Properties}

\begin{figure}
\includegraphics[width=8.5cm]{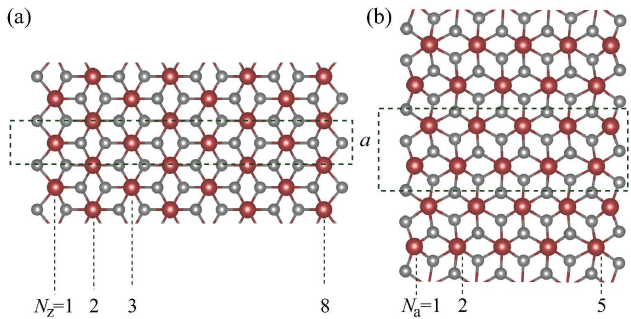}
\caption{\label{zigzagarm}
(Color online) Top view of (a) zigzag and (b) armchair TiSe$_2$ nanoribbons. The unitcell is indicated by the dashed 
box.}
\end{figure}

The TiSe$_2$ nanoribbons (TiSe$_2$-NRs) are obtained by cutting the $2$D-TiSe$_2$ monolayer. According to the different 
directions of termination, there are two  kinds of 
nanoribbons: zigzag (TiSe$_2$-ZNR), and armchair (TiSe$_2$-ANR). Apart from the termination, TiSe$_2$-NRs are defined 
by their widths. The width of the zigzag nanoribbon is denoted as $N_z$ (TiSe$_2$-$N_z$ZNR) and 
for armchair nanoribbon, the width is denoted by $N_a$ (TiSe$_2$-$N_a$ANR). In Fig. 
\ref{zigzagarm} the lattice structure of TiSe$_2$-8ZNR and TiSe$_2$-5ANR are presented. 
In our calculations, we consider width $N_z$ from 2 to 10 and  $N_a$  from 2 to 8.

The fully optimized NRs exhibit structural deviation at the edges. For example TiSe$_2$-ANRs are strongly distorted after relaxation, compared to TiSe$_2$-ZNRs. In the triple layer networks, the edge selenium atoms shift their position from the Se layers to the Ti layer for both zigzag and armchair nanoribbons whereas the Ti atoms at the edges shift their position from the Ti layer to the Se layers for only zigzag nanoribbons. At one of the edges the Ti atom is closer to the lower Se layer, and  the Ti atom at the other edge is closer to the upper Se layer. As  seen in Fig. \ref{zigzagarm}(b) for armchair nanoribbons reconstruction takes place, as the Ti atoms at the edges moved towards the ribbon's center and the Se atoms tend to shift slightly outward. For TiSe$_2$-8ZNR, shown in Fig. 
\ref{zigzagarm}(a), the Ti atoms moved slightly out of the plane, leading to a change of the Ti-Se bond length along 
the ribbon-axis. Nevertheless, the triple-layer  networks are well kept intact for both ribbons. For instance, the 
average 
Ti-Se bond lengths for TiSe$_2$-7ZNR are $2.56$\AA{} in the inner site, and $2.44$\AA{} at the two edges. The angle 
between  Se-Ti-Se bond is $6.22^{\circ}$ between the center and edge of the $N_z$=7 zigzag nanoribbon. For 
the TiSe$_2$-8ANR, coordination of atoms are different so that the Ti-Se bond length is different with values of 
$2.50$, $2.57$, and $2.64$ \AA{} in the inner site, at the edges it  decreases to $2.38$\AA{}. All of the nanoribbons 
display the same structural property, and the only difference is that the bond lengths between the edge Ti-Se atoms are 
longer in ZNRs than those in ANRs. Similar to the case of MoS$_2$ nanoribbons\cite{li-mos2nr}, at the edges the Ti-Se 
bond lengths decrease because of the irregular force on the edge atoms. Also, a Bader charge analysis tells us that 
charges on both Ti and Se atoms are equally distributed along the ribbon axis, since all of the Ti atoms lose  
the same amount of electron charge which is taken by the Se atoms. Likewise in the 2D-TiSe$_2$ layer, every Ti atom 
loses 1.4 
electrons to the Se atoms which gain 0.7 electrons along the ribbon axis.

\begin{figure}[htbp]
\includegraphics[width=8.5cm]{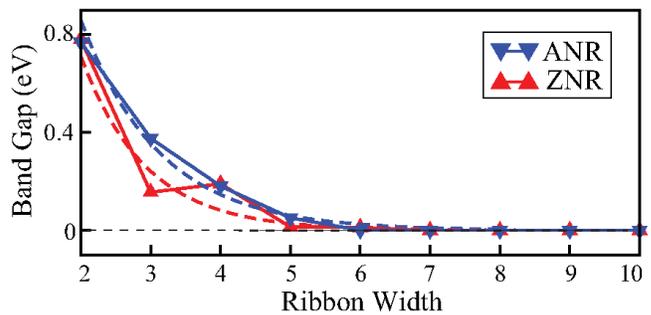}
\caption{\label{gapvswidth}
(Color online) Energy gap of zigzag ($2 \leq N_{z}\leq10$) and armchair ($2 \leq 
N_{a}\leq10$) 1T-TiSe$_2$ nanoribbons as 
function of the ribbon width. Dashed curves are exponential fits.}
\end{figure}

\subsection{Electronic Properties}

\begin{figure}
\includegraphics[width=8.5cm]{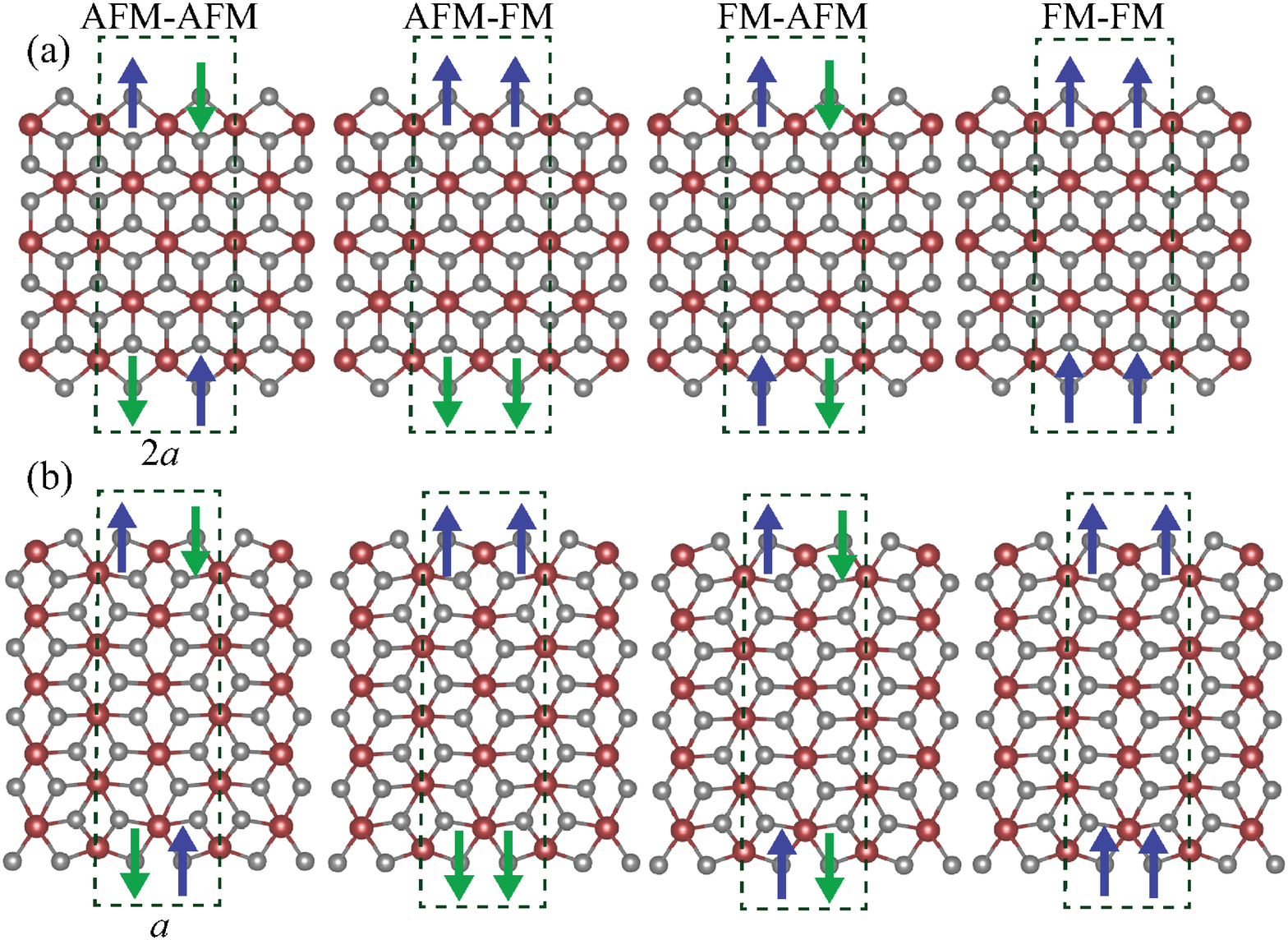}
\caption{\label{mag}
(Color online) Different magnetic interaction cases for (a) TiSe$_2$-5ZNR and (b) TiSe$_2$-5ANR.}
\end{figure}

During the geometry optimization, we first carried out both spin-polarized and spin-unpolarized total energy 
calculations in order to determine the ground state of the different TiSe$_2$-$N_z$ZNR 
(TiSe$_2$-$N_a$ANR). There is no 
energy difference between spin-polarized and  spin-unpolarized calculations which indicates that zigzag and armchair 
TiSe$_2$ nanoribbons have a nonmagnetic ground state. To be more confident about the magnetization of the edges, we 
also 
performed calculations for four different magnetic orderings for TiSe$_2$-4ZNR and also TiSe$_2$-5ZNR   by taking a 
double 
unitcell, such as antiferromagnetic (AFM), ferromagnetic (FM) (where, the atoms are located at different edges are AFM 
coupled, and at the same edge are FM coupled) (see Fig. \ref{mag}(a)).  We take the case of a TiSe$_2$-5ZNR as an 
example. 
Calculations starting from the four magnetic states, namely AFM-AFM, AFM-FM, FM-AFM, and FM-FM, and results in the same 
total energy. The same magnetic test is also applied to armchair nanoribbons (see Fig. \ref{mag}(b)). All the test 
results gave the  same total energy and zero net magnetic moment. As a result, TiSe$_2$ armchair nanoribbons have a  
nonmagnetic ground state like MoS$_2$-ANRs\cite{li-mos2nr}. Thus, our calculation demonstrates that TiSe$_2$-ZNRs and 
TiSe$_2$-ANRs are not magnetic and the edge states do not effect the magnetization of  the structures.

After analyzing the structural and magnetic properties, we investigated the band dispersion of the TiSe$_2$-NRs. 
Electronic structures of TiSe$_2$-NRs show similar behavior like the single-layer 1T-TiSe$_2$. In fact, we 
found that reducing 
the dimensionality from 2D to 1D, at a certain ribbon width a metal to semiconductor transition is found for 
both 
zigzag and armchair nanoribbons as seen in Fig. \ref{gapvswidth}. The band gap decays monotonically with the ribbon 
width for armchair 
nanoribbons, however for zigzag nanoribbons the rapid band gap decrease is superposed with an even-odd oscillation with 
increasing $N_z$ and finally both 
structures switches to the zero energy gap of monolayer TiSe$_2$ (for $N_z$ $\geq7$, and $N_a$ 
$\geq6$). 
Similar oscillatory behavior is  
also observed in the equilibrium lattice constant for TiSe$_2$-$N_z$ZNRs, when we increase the ribbon width 
$N_z$, the 
lattice constant approached slowly the value $3.52$\AA{} which is the same as that calculated for the $2$D-TiSe$_2$. 
The 
edge reconstructions are more effective in changing the equilibrium lattice constant of ultra narrow 
ribbons. 

As illustrated in Fig. \ref{gapvswidth}, the band gaps as a function of ribbon width for 
both zigzag and armchair-edged nanoribbons decay very rapidly, except for a small superposed oscillation observed in 
ultranarrow zigzag nanoribbons. Similar band gap oscillations as a function of ribbon width  
were also predicted for other semiconducting nanoribbons\cite{son-gnr}. Nevertheless, due to the rapid decay in both 
types of nanoribbons, to provide a quantitative measure for these decays the band gap variations are fitted to the 
exponential functions, E$_{gap}$(N)= $\alpha$ 
exp(-N$\beta$), where N is the width of the nanoribbon (for ZNRs N=$N_z$ and for ANRs N=$N_a$), 
and $\alpha$ and $\beta$ are fitting parameters.  For armchair and zigzag nanoribbons, the values of the fitting 
parameters are found to be 
$\alpha$=$5.06$, $\beta$=$0.89$ eV and $\alpha$=$6.17$, $\beta$=$1.08$ eV, respectively. For
$N\geq7$, both types of nanoribbons show metallic behavior.

\begin{figure}
\includegraphics[width=8.5cm]{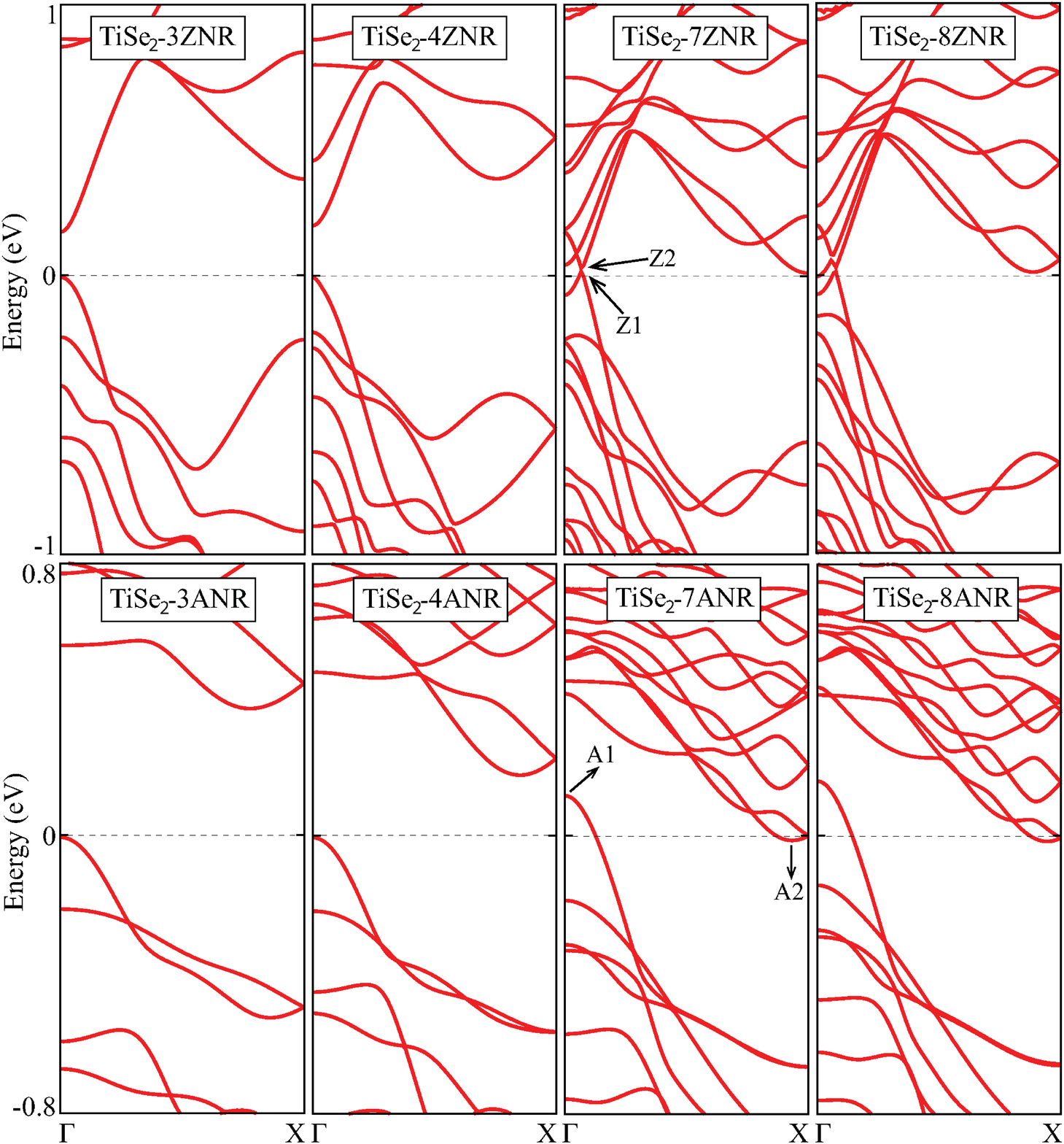}
\caption{\label{bands}
(Color online) Electronic band structure of a series of zigzag and armchair nanoribbons of 1T-TiSe$_2$ by using the PBE 
method.}
\end{figure}

Spin-unpolarized band structures of TiSe$_2$-$N_z$ZNRs are presented in Fig. \ref{bands}. Notice that 
the band 
structures show similar property at the X-point for odd and even numbers of ribbon width. For the ribbon width 
of $N_z$=2 a large gap of about $0.786$ eV is found. Among the four ZNRs in Fig. \ref{bands}, 
TiSe$_2$-4ZNR has 
the largest band gap of $0.201$ eV, TiSe$_2$-3ZNR has a medium band gap of $0.165$ eV, TiSe$_2$-5ZNR and 
TiSe$_2$-6ZNR 
have the smallest band gaps of $12$ meV and $5$ meV, respectively. Both the direct band 
gap of zigzag-edged and the indirect band gap of armchair-edged ultranarrow nanoribbons decrease with increasing 
ribbon width and eventually vanish for $N_z$ $\geq7$, and $N_a$ 
$\geq6$. The conduction band minimum (CBM) and the valance band 
maximum (VBM) cross resulting in a semimetallic band structure with overlapping bands. 

\begin{figure}
\includegraphics[width=8.5cm]{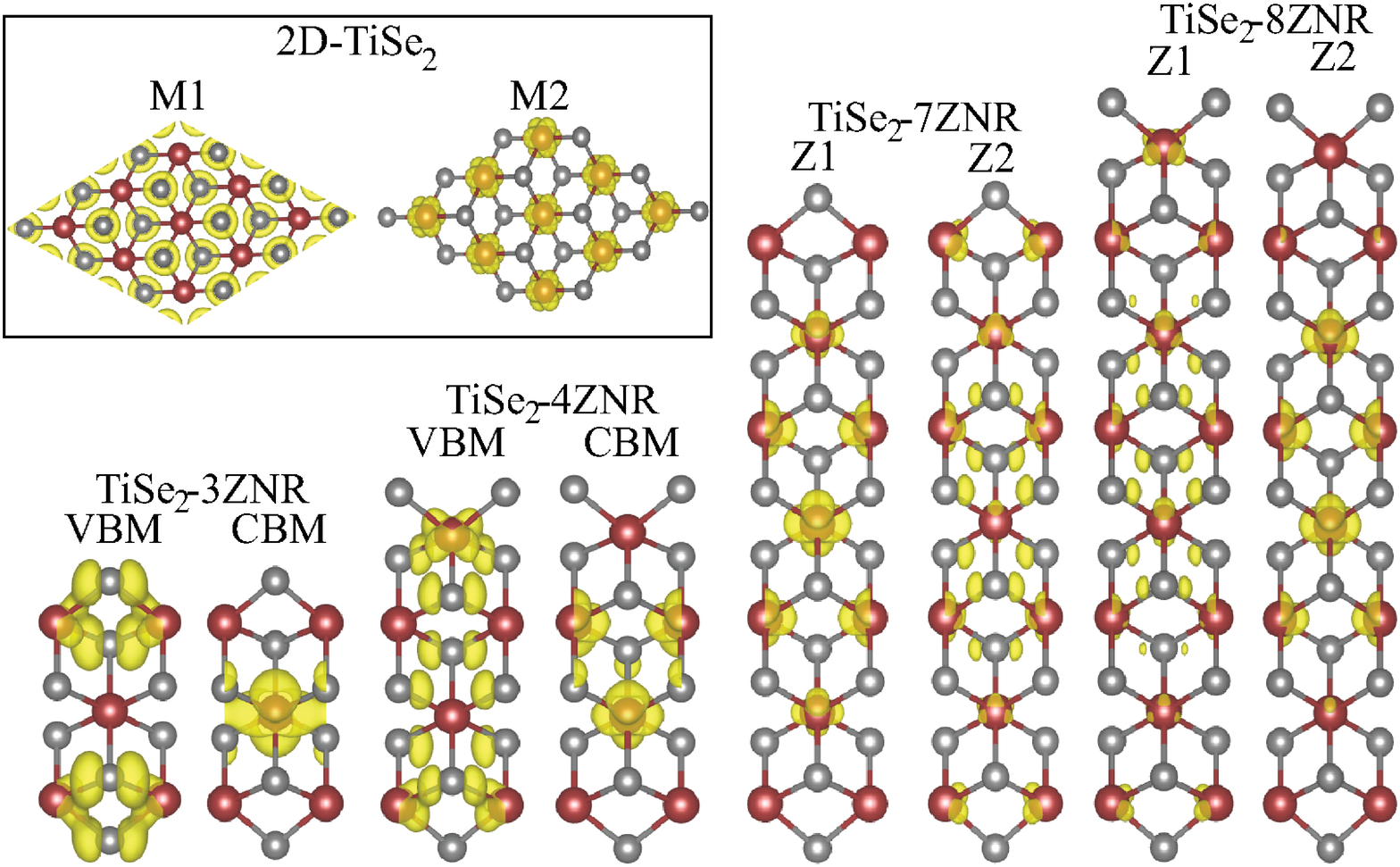}
\caption{\label{zigzagband-dec}
(Color online) Band decomposed charge density plots of monolayer and $N_z$=3,4,7,8 nanoribbons of TiSe$_2$ 
where Z1 and Z2 are shown in the band-structures (see Fig. \ref{bands}). Inset shows the $\Gamma$-point charge 
densities of M1 and M2 band edges (shown in Fig. \ref{bandmonotise2}) of 2D TiSe$_2$.}
\end{figure}

In order to investigate this 
width-dependent transition in the band structure, as well as the odd-even variations observed in the narrowest ZNRs, we 
have considered partial charge density (PCD) profiles corresponding to VBM and CBM, or for some specific pair of points 
in the band structures. These pair of points are M1 and M2 for 2D-TiSe2 (Fig. \ref{bandmonotise2}), Z1 and Z2 for ZNRs, 
and A1 and A2 for ANRs (Fig. \ref{bands}). The PCD plots of the VBM and the CBM  as shown in Fig. \ref{zigzagband-dec} 
indicate the electronic states around the Fermi level. 
For TiSe$_2$-3ZNR (TiSe$_2$-4ZNR), the VBM and the CBM originate from a hybridized mixture of 3d electrons of Ti and 4p 
electrons of Se atoms with the hybridization being stronger in the VBM than that in the CBM. A comparison of the VBM 
states of TiSe$_2$-3ZNR and TiSe$_2$-4ZNR indicate that they are localized more at the edges for odd $N_z$, 
whereas they are more uniform distributed for even $N_z$ ribbons. For wider ribbons ($N_z$ $>$ 
$4$), both 
the VBM and CBM states tend to delocalize and the metallic character is attained (this is evident for 
$N_z$=7 and  $N_z$=8 in Fig. \ref{zigzagband-dec}). With increasing $N_z$, the PCD 
plots at the Z1 and Z2 points tend to converge to those at the M1 and M2 pair for 2D-TiSe$_2$, where the corresponding 
states are localized on 
the Se and Ti atoms, respectively. The opening of a band gap in very narrow ribbons can be attributed to quantum size 
effects.

\begin{figure}
\includegraphics[width=8.5cm]{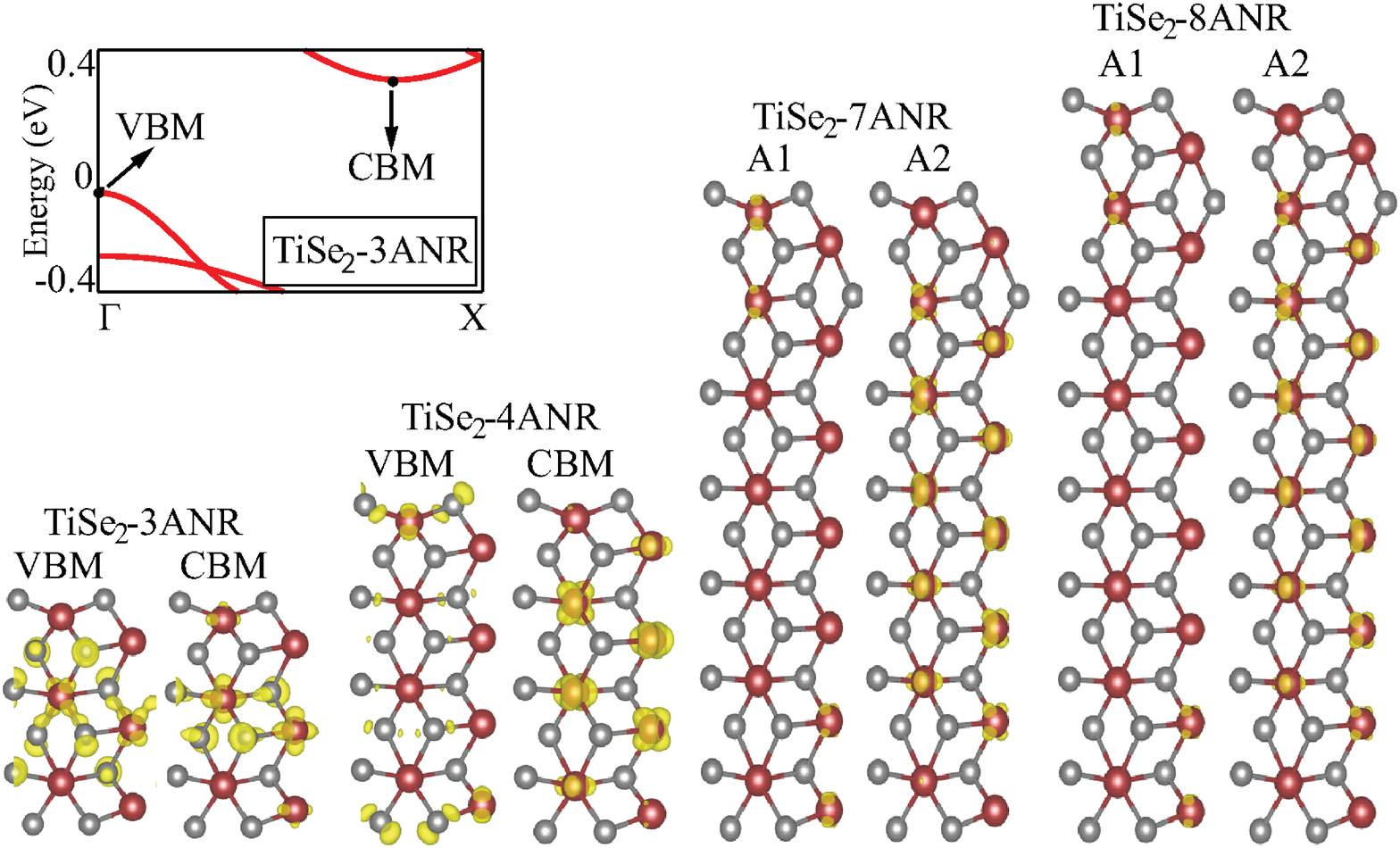}
\caption{\label{armband-dec}
(Color online) TiSe$_2$-3ANR band structure and band decomposed charge densities of $N_a$=3,4,7,8 
nanoribbons of TiSe$_2$ where the A1 and A2 refer to the states indicated in Fig. \ref{bands}.}
\end{figure}

Typical band structures for a series of armchair TiSe$_2$ nanoribbons are also shown  in Fig. \ref{bands}. 
Unlike zigzag nanoribbons, the electronic structure of the armchair ribbons exhibit an indirect band gap for 
$N_a$ $\leq$6. The gap decreases exponentially with the ribbon width. The band gap is almost halved 
when the ribbon width is increased from $N_a$=2 to $N_a$=4. TiSe$_2$-$5$ANR
still has a band gap of about $5.2$ meV. Starting with $N_a$=6, the CBM dips into the Fermi level, so that 
the armchair nanoribbons become metallic for wider widths. Some partial 
charge density plots for TiSe$_2$-$N_a$ANRs are also illustrated in Fig. \ref{armband-dec}. Similarly, the 
VBM and CBM states are composed 
of a hybridized mixture of Ti-3d and Se-4p orbitals for small nanoribbons, however for the ribbon width larger than 
four, the hybridization becomes lost. 

\section{Hydrogen Termination of Edges}\label{hpass}

In order to investigate the effect of dangling states present at the edges of the nanoribbons, we have passivated 
the edge atoms by hydrogen atoms. These unsaturated bonds influence the electronic properties of the ribbons. 
Naturally these 
states do 
not exist in the infinite TiSe$_2$ single layer, therefore reducing dimensionality from 2D to 1D it will be 
of importance control the dangling bonds. Earlier, it was shown for graphene nanoribbons that when the dangling 
bonds at the 
edges are passivated with hydrogen atoms the electronic and magnetic properties of the ribbons are 
modified.\cite{veronica-gnr}  Unlike graphene, the TiSe$_2$-NRs have two types of atoms at the edges so that 
both Ti and Se atoms have to be passivated by hydrogen atoms to compensate  the edge states. 

\begin{figure}[htbp]
\includegraphics[width=8.5cm]{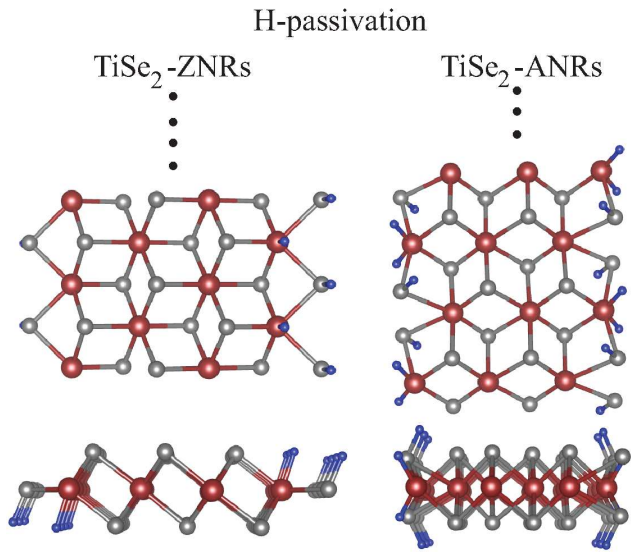}
\caption{\label{hpas-types}
(Color online) Passivation of the edge states with hydrogen atoms (blue colored) 
for the zigzag and armchair nanoribbons.}
\end{figure}

\begin{figure}[htbp]
\includegraphics[width=8.5cm]{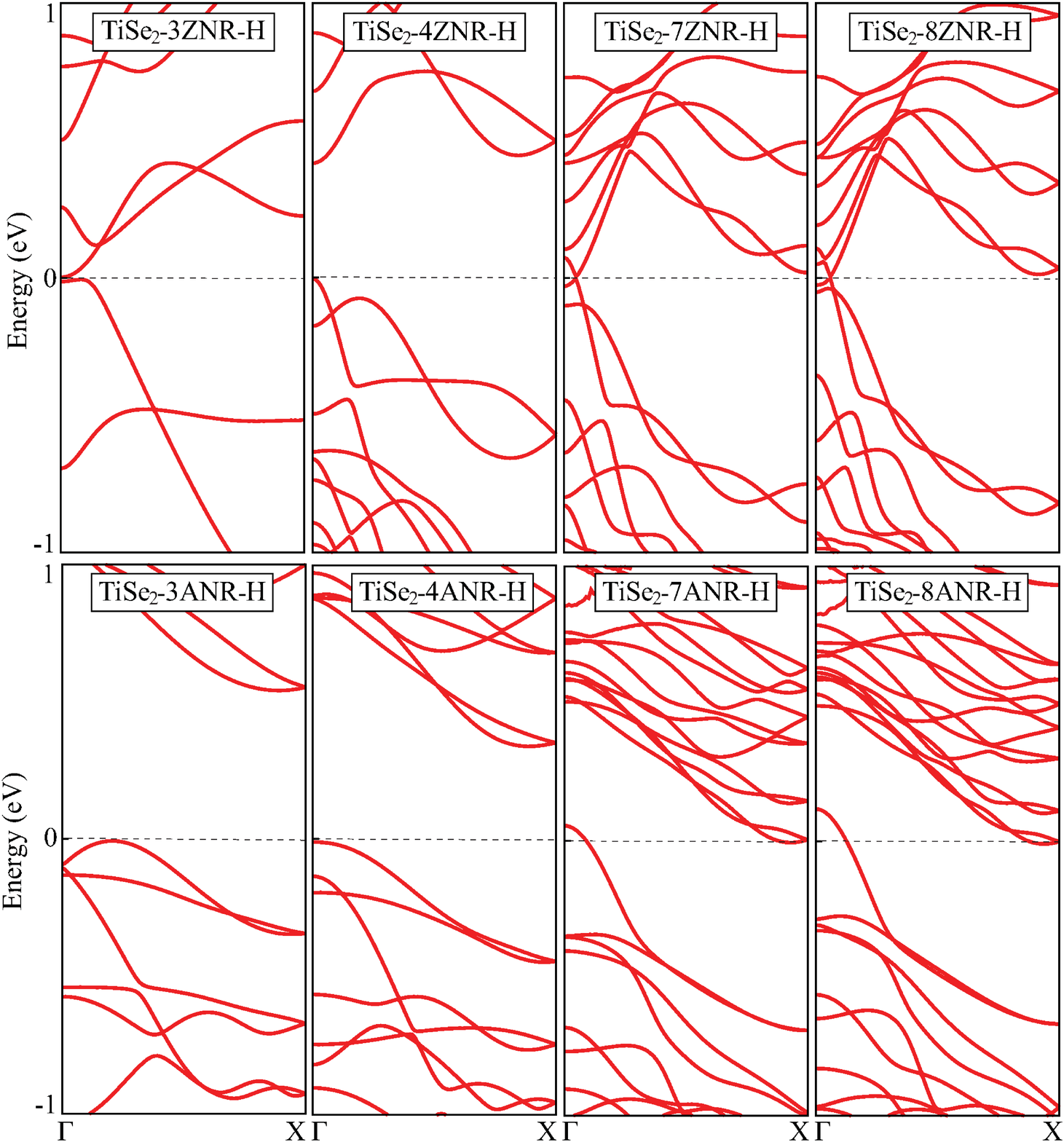}
\caption{\label{band-hpas}
(Color online) Band structures for zigzag and armchair nanoribbons where the 
edge atoms are passivated by hydrogen 
atoms.}
\end{figure}

Among possible configurations for the edge termination with hydrogen atoms, the most energetically 
favorable structure is shown for the TiSe$_2$-4ZNR in Fig. \ref{hpas-types}. As seen in the figure
where the edge atoms are passivated by hydrogen atoms symmetrically, hydrogenation of the nanoribbons also enhances the 
stability of the structures. After hydrogenation the ground state energies is lowered, and the binding 
energy is found to be $11.7$ eV for the case of TiSe$_2$-4ZNR. The band structures for several hydrogenated ZNRs are 
shown in 
Fig. \ref{band-hpas}. The TiSe$_2$-$N_z$ZNRs are all metallic except for $N_z$=4.

We performed a analysis for the armchair nanoribbons. In TiSe$_2$-3ANR, as an example shown in Fig. 
\ref{hpas-types}, the edge Se and Ti atoms 
are passivated by one and two hydrogen atoms, respectively. 
The binding energy of the TiSe$_2$-3ANR is 23.4 eV. After the hydogenation, TiSe$_2$-3ANR and TiSe$_2$-4ANR are 
semiconductors with an increasing band gap. Also, the VBM state moves a little away from the 
$\Gamma$-point in case of $N_a$=3. TiSe$_2$-7ANR and  TiSe$_2$-8ANR are still metallic after 
hydrogenation, 
however the overlap of the conduction and valance bands is reduced.

\section{Conclusions}\label{conc}

In this work, we have investigated the electronic and magnetic properties of zigzag 
and armchair-edged TiSe$_2$ nanoribbons by means of first-principles 
calculations. Overall, our results demonstrate that
these TMD nanoribbons which are in 1T phase have quite different
characteristics from nanoribbons of other widely studied materials such as graphene or MoS$_2$. Our calculations 
revealed that only ultranarrow zigzag 
and armchair nanoribbons exhibit semiconducting behavior and their band gap 
rapidly decreases to zero with increasing ribbon width. $N_a$ $\geq$ 6 and $N_z$ $\geq$ 7 
nanoribbons exhibit metallic behavior like two-dimensional TiSe$_2$. The width dependency of the band 
gap can be fairly represented by an exponential decay function. Both zigzag and armchair ribbons have 
nonmagnetic ground states. In addition, the robust metallic behavior of both zigzag and armchair TiSe$_2$ nanoribbons 
remains unaltered even after passivation of the edges by hydrogen atoms. The metallic character of the 
wider ribbons of TiSe$_2$ regardless of their edge symmetry is an advantageous property for utilizing them as 
one-dimensional interconnects of nanoscale circuits.

\begin{acknowledgments}

This work was supported by the Flemish Science Foundation (FWO-Vl) and the 
Methusalem foundation of the Flemish government. Computational resources were 
provided by TUBITAK ULAKBIM, High Performance and Grid Computing Center 
(TR-Grid e-Infrastructure). HS is supported by a FWO Pegasus Long Marie Curie 
Fellowship. JK is supported by a FWO Pegasus Short Marie Curie Fellowship. 
HDO, HS and RTS acknowledge the support from TUBITAK through project 
114F397.  
\end{acknowledgments}

\end{document}